\documentstyle[preprint,prd,aps]{revtex}
\begin{document}
\draft

\title {O(d, d)--Symmetry and Ernst  Formulation for\\
Einstein--Kalb--Ramond Theory in Two Dimensions}

\author{Alfredo Herrera}

\address{Joint Institute for Nuclear Research,\\
Dubna, Moscow Region 141980, RUSSIA, \\
e-mail: alfa@cv.jinr.dubna.su}

\author {\rm and}

\author{Oleg Kechkin}

\address{Nuclear Physics Institute,\\
Moscow State University, \\
Moscow 119899, RUSSIA, \\
e-mail: kechkin@cdfe.npi.msu.su}

\date{December 1996}

\maketitle

\draft
\begin{abstract}
The (3+d)--dimensional Einstein--Kalb--Ramond theory reduced to two
dimensions is considered. It is shown that the theory allows two different
Ernst--like $d \times d$ matrix formulations: the real non--dualized target
space and the Hermitian dualized non--target space ones. The $O(d, d)$
symmetry is written in a $SL(2,R)$ matrix--valued form in both cases.
The Kramer--Neugebauer transformation, which algebraically maps the
non--dualized Ernst potential into the dualized one, is presented.
\end{abstract}

\draft

\narrowtext
\section{Introduction}
Heterotic string theory predicts the effective action for its
massless excitations.
The bosonic sector of this action provides the
correct description of the  gravitational, Kalb--Ramond, dilaton and
gauge vector fields at Plank scales. The continuous symmetries of
the effective system correspond to the discrete ones for the exact
string theory \cite {s1}--\cite {s2}.

In the previous paper \cite {hk} we have shown that the simplified
string gravity system, the Einstein--Kalb--Ramond (EKR) theory, being
reduced to three dimensions, allows two formulations which are very similar
to the pure Einstein theory. The first of them
corresponds to the (target space) Ernst formulation of the stationary
Einstein equations \cite {e}, while the second one is related to their
(non--target space) metric representation.

Here we discuss the properties of EKR theory reduced to two dimensions. We
show that one can introduce a new Hermitian Ernst--like matrix potential
in the non--target space formulation. It is established that this
potential can be mapped into the real target space Ernst--like matrix one using
a complex transformation which generalizes the Kramer--Neugebauer map
for the stationary axisymmetric Einstein theory \cite {kn}.

It is shown that
the real and imaginary parts of the new Ernst--like potential define a
non--coset chiral matrix which possesses the properties of the
Belinsky--Zakharov one for the vacuum system \cite {bz}.
This new matrix provides the explicit $O(d, d)$--invariant representation of
the theory under consideration.
The global $O(d, d)$ symmetry transformations are rewritten in
the matrix--valued $SL(2, R)$ form using both Ernst--like matrix potentials.
\section{Ernst Matrix Potentials}
The action for EKR theory in $(3 + d)$ dimensions is:
\begin{equation}
{\cal S} = \int d^{3+d}x {\mid {\cal G} \mid}^{\frac {1}{2}} \left\{ - {\cal R} +
\frac {1}{12} {\cal H}^2 \right\},
\end{equation}
where ${\cal R}$ is the Ricci scalar for the metric ${\cal G}_{M N}$,
$(M = 0, ..., 3 + d)$
and
\begin{equation}
{\cal H}_{MNL} = \partial _{M} {\cal B}_{NL}
+ {\rm cyc.\,\, perms.}
\end{equation}
Here ${\cal B}_{NL}$ is the antisymmetric Kalb--Ramond field and
${\cal H}_{MNL}$ is the non--dualized axion one.

As in the previous work \cite {hk} we consider the ansatz
\begin{equation}
{\cal G}_{\mu, n + 2} = {\cal B}_{\mu, n + 2} = 0,
\end{equation}
where $\mu = 0, 1, 2; \, n = 1, ..., d$. Following \cite {ms} we perform the
compactification of $d$ dimensions on a torus. The resulting theory is
a $\sigma$--model constructed on the matrix fields
$G_{mn} = {\cal G}_{m + 2, n + 2}$ and $B_{mn} = {\cal B}_{m + 2, n + 2}$
coupled to 3--gravity with the metric $g_{\mu \nu} = {\cal G}_{\mu \nu}$.

In this paper we deal with time--independent field configurations, when
the 3--metric can be parametrized in the Lewis--Papapetrou form:
\begin{equation}
ds_3^2 = e^{2\gamma}\left (d\rho ^2 + dz^2 \right ) - \rho ^2d\tau ^2.
\end{equation}
Then the ``material part'' of the motion equations is
\begin{equation}
\nabla (\rho J^B) - \rho J^GJ^B = 0,
\end{equation}
\begin{equation}
\nabla (\rho J^G) - \rho (J^B)^2 = 0,
\end{equation}
where $J^B = \nabla B \, G^{-1}$, $J^G = \nabla G \, G^{-1}$ and the operator
$\nabla = \{\partial _{\rho}, \partial _z\}$. These equations are the
Euler--Lagrange ones for the effective 2--dimensional action
\begin{equation}
^2S = \frac {1}{4}\int d\rho dz \rho Tr\left [(J^G)^2 - (J^B)^2 \right ].
\end{equation}

The defining function $\gamma$ relations can be obtained from the
3--dimensional Einstein equations \cite {hk}; as result we have
\begin{eqnarray}
\gamma _{,z} &=& \frac {1}{4}\rho Tr\left [ J^G_{\rho}J^G_{z} -
J^B_{\rho}J^B_{z} \right ],
\nonumber \\
\gamma _{,\rho} &=& \frac {1}{8}\rho Tr\left \{\left [(J^G_{\rho})^2 -
(J^G_{z})^2 \right ] -
\left [(J^B_{\rho})^2 - (J^B_{z})^2\right ]\right \}.
\end{eqnarray}
Eq. (5), being rewritten in the form
$\nabla [\rho G^{-1}(\nabla B)G^{-1}] = 0$, becomes the compatibility
condition for the relation defining the antisymmetric matrix $\Omega$:
\begin{equation}
\nabla \Omega = \rho G^{-1}(\tilde \nabla B)G^{-1}.
\end{equation}
(Here $\tilde \nabla _{\rho } = \nabla _z$ and $\tilde \nabla _z =
- \nabla _{\rho}$), see \cite {k}.) This new matrix $\Omega$, together with the
original one $G$, provide an alternative Lagrange description of the problem.
Namely, from Eq. (9) one obtains that $\nabla [\rho G(\nabla \Omega)G] = 0$,
i.e.,
\begin{equation}
\nabla \left (\rho ^{-1}J^{\Omega} \right ) + \rho ^{-1}J^{\Omega}J^G = 0,
\end{equation}
where $J^{\Omega} = G \, \nabla \Omega$.
This equation, together with Eq. (6) rewritten as
\begin{equation}
\nabla (\rho J^G) - \rho ^{-1}(J^{\Omega})^2 = 0,
\end{equation}
constitute the Euler--Lagrange system for the action
\begin{equation}
^2S = \frac {1}{4}\int d\rho dz Tr\left [\rho (J^G)^2 + \rho ^{-1} (J^{\Omega})^2 \right ].
\end{equation}

In \cite {hk} it was shown that the alternative Lagrange formulation
in three dimensions is
connected with the use of the antisymmetric vector matrix $\vec \Omega$. It is
easy to see that in two dimensions the only $\tau$--component is
non--trivial and $\Omega = (\vec \Omega)_{\tau}$.

Using Eq. (9) one can also represent the relations (8) defining the function
$\gamma$ in terms of the matrices $G$ and $\Omega$:
\begin{eqnarray}
\gamma _{,z} &=& \frac {1}{4}Tr\left [\rho J^G_{\rho}J^G_{z} +
\rho ^{-1}J^{\Omega}_{\rho}J^{\Omega}_{z} \right ],
\nonumber \\
\gamma _{,\rho} &=& \frac {1}{8}Tr\left \{\rho \left [(J^G_{\rho})^2 -
(J^G_{z})^2 \right ] +
\rho ^{-1}\left [(J^{\Omega}_{\rho})^2 - (J^{\Omega}_{z})^2\right ]\right \}.
\end{eqnarray}

In \cite {hk} the formal analogy between the
EKR theory, on the one hand,  and the Einstein and
Einstein--Maxwell--Dilaton--Axion theories, on the other hand,
(and moreover, between EKR and an arbitrary symplectic gravity
model \cite {ky1}) was established. Using this analogy, one can suppose
the existence of the algebraical transformation which directly maps
the $(G, \, B)$--formalism into the $(G, \, \Omega)$ one. It is
easy to check that the complex transformation
\begin{equation}
G \rightarrow \rho G^{-1}, \qquad B \rightarrow i\Omega,
\end{equation}
actually maps Eqs. (5)--(6) into the Eqs. (10)--(11). Thus, the EKR system
allows
a Kramer--Neugebauer--like transformation \cite {kn} in two dimensions.
Using the relations (13), defining the function $\gamma$, one can see
that this function undergoes the non--trivial transformation
\begin{equation}
e^{2\gamma} \rightarrow
\frac {\rho ^{\frac {d}{4}}}{{\mid det G\mid}^{\frac {1}{2}}}e^{2\gamma}.
\end{equation}
under the map (14).
The Kramer--Neugebauer--like transformation for the
Einstein--Maxwell--Dilaton--Axion theory was found in \cite {ky2}, and
for the general case of symplectic models with the coset space $Sp(2n, R)/U(n)$,
in \cite {ky1}.

In the previous work \cite {hk} it was shown that the $d \times d$ matrix
variable
\begin{equation}
X = G + B,
\end{equation}
which was firstly entered in \cite {ms}, realizes the real Ernst--like
formulation of the problem in three dimensions. It is easy to prove
that the 2--dimensional action (7) can be rewritten as
\begin{equation}
^2S = \frac {1}{2}\int d\rho dz \rho Tr\left [J^X J^{X^T} \right ],
\end{equation}
where $J^X = \nabla X \, (X + X^T)^{-1}$. A remarkable fact is that
in two dimensions one can introduce the new Ernst--like
matrix potential
\begin{equation}
E = E^+ = \rho G^{-1} + i\Omega.
\end{equation}
The use of $E$ allows to represent the non--target space action (12) in
a similar to Eq. (17) form
\begin{equation}
^2S = \frac {1}{2}\int d\rho dz \rho Tr\left [J^E J^{\bar E} \right ],
\end{equation}
where $J^E = \nabla E \, (E + \bar E)^{-1} - \frac {1}{2} \nabla ln \rho$.

The introduced above Ernst--like potentials $X$ and $E$ allow
to extremely simplify the form of the Kramer--Neugebauer
transformation (14):
\begin{equation}
X \rightarrow E.
\end{equation}
In the next section we will show that these matrices provide a natural
language for the symmetry analysis of the theory under
consideration.
\section{O(d, d)--symmetry in SL(2, R) Form}
In \cite {ms} the above discussed matrices $G$ and $B$ had been combined into
the $2d \times 2d$ matrix
\begin{eqnarray}
M = \left (\begin{array}{crc}
G^{-1} &\quad & - G^{-1} B\\
B G^{-1} &\quad & G - BG^{-1}G\\
\end{array}\right ),
\end{eqnarray}
which allows to transform the action (7) into the chiral form
\begin{equation}
^2S = \frac {1}{8}\int d\rho dz \rho Tr\left [(J^M)^2 \right],
\end{equation}
where $J^M = \nabla M M^{-1}$.

It is easy to check that $M^T = M$ and
\begin{eqnarray}
M^T \eta M = \eta, \qquad {\rm where} \qquad
\eta = \left (\begin{array}{crc}
0 & I\\
I & 0\\
\end{array}\right ),
\end{eqnarray}
i.e., the null--curvature matrix $M$ belongs to the coset
$O(d, d)/O(d) \times O(d)$ \cite {ms}. The isometry transformations
\begin{equation}
M \rightarrow C^T M C,
\end{equation}
which preserve the action (22), are defined by an arbitrary matrix
$C$ belonging to the $O(d, d)$ group. In \cite {hk} it was shown that
the Gauss decomposition
\begin{eqnarray}
C = \left (\begin{array}{crc}
(S^T)^{-1} & \quad & - (S^T)^{-1}R\\
- L(S^T)^{-1} & \quad & S + L(S^T)^{-1}R\\
\end{array}\right ),
\end{eqnarray}
where $R^T = - R$ and $L^T = - L$, defines that part of the
$O(d, d)$ group  which can be continuously transformed into the
unit element.

Also in \cite {hk} it was established that, in terms of the matrices
$G$ and $B$, the transformation (24) can be written in a matrix--valued
$SL(2, R)$ form:
\begin{equation}
X \rightarrow S^T(X^{-1} + L)^{-1}S + R.
\end{equation}

Now it is natural to ask how this transformation acts on the matrix
variables $G$ and $\Omega$ which provide an alternative description of
the problem. To answer this question one can introduce the new
$2d \times 2d$ matrix $N$
\begin{eqnarray}
N = \left (\begin{array}{ccc}
G & \quad & - G\Omega\\
\Omega G & \quad & - \Omega G \Omega - \rho ^2 G^{-1}\\
\end{array}\right ),
\end{eqnarray}
which is symmetric and satisfies the nongroup relation
\begin{equation}
N \eta N = - \rho ^2 \eta,
\end{equation}
instead of the group one (23).
A similar situation arises in the stationary
axisymmetric case of the pure Einstein theory \cite {bz}.

The 2--dimensional action (12) in terms of $N$,
up to a bound term, has the same form as the
action (22):
\begin{equation}
^2S = \frac {1}{8}\int d\rho dz \rho Tr\left [(J^N)^2 \right],
\end{equation}
where $J^N = \nabla N \, N^{-1}$.\,
It is easy to see that
the relation (28) preserves under the transformation
\begin{equation}
N \rightarrow C^T N C,
\end{equation}
where $C$ is an arbitrary matrix belonging to the $O(d, d)$ group.

Thus, this transformation constitutes a symmetry for the action (29).
However, it has no the sense of isometry transformation since the
matrix variable $\Omega$ is not a target space potential. By
analogy with the stationary axisymmetric Einstein system,
this $O(d, d)$ symmetry representation can be compared
with the $SL(2, R)$ one of the non--dualized Einstein equations.

Therefore,
performing a straightforward calculation, one can establish that the
map (30) in terms of the Ernst--like matrix potential $E$ also takes
a $SL(2, R)$ matrix--valued form:
\begin{equation}
E \rightarrow S^T(E^{-1} - iL)^{-1}S + iR.
\end{equation}
Thus, both
Ernst--like potentials $X$ and $E$ transform in a similar way which
is a direct generalization of the $SL(2, R)$ transformation for the
vacuum theory.

At the end of the paper it must be noted that there are non--trivial $O(d, d)$
transformations which can not be represented in the
Gauss decomposition form (27).
As an example one can take the transformation defined by the matrix $C = \eta$,
which is an $O(d, d)$ one and corresponds to the map $M \rightarrow M^{-1}$ (or
$N \rightarrow -\rho ^2 N^{-1}$, correspondingly). Such a kind of
transformations was firstly discussed in \cite {s2} as ``strong--weak
coupling duality'' in three dimensions. Using Ernst--like potentials, this
transformation can be rewritten as
\begin{equation}
X \rightarrow X^{-1}; \qquad E \rightarrow E^{-1}.
\end{equation}

\section{Conclusion}
The existence of the matrix $N$ allows to construct the
Hauser--Ernst--like linear system \cite {he} and to derive the
infinite Geroch--like group \cite {g}. This group will be the loop
expansion of the isometry group $O(d, d)$ discussed here. In the particular
case of $d=2$ this procedure was performed in \cite {b}.
\acknowledgements
We would like to thank our colleagues from the JINR and NPI
for an encouraging relation to our work. One of the authors (A. H.) would
like to thank CONACYT and SEP for partial financial support.

\end{document}